# Willingness to Pay to Prevent Water and Sanitation-Related Diseases Suffered by Slum Dwellers and Beneficiary Households: Evidence from Chittagong, Bangladesh

Mohammad Nur Nobi*

## Abstract:

*A significant proportion of slum residents offer vital services that are relied upon by wealthier urban residents. However, the lack of access to clean drinking water and adequate sanitation facilities causes considerable health risks for slum residents, leading to interruption in services and potential transmission of diseases to the beneficiaries. This study explores the willingness of the households benefitting from these services to contribute financially towards the measures that can mitigate the negative externalities of the diseases resulting from poor water and sanitation in slums. This study adopts the Contingent Valuation Method using face-to-face interviews with 260 service-receiving households in Chittagong City Corporation of Bangladesh. Estimating the logistic regression model, the findings indicate that 74 percent of respondents express their willingness to contribute financially towards an improvement of water and sanitation facilities in the slums. Within this group, 16 percent are willing to pay 1.88 USD/month, 18 percent prefer 3.86 USD/year, and 40 percent are willing to contribute a lump sum of 3.92 USD. The empirical findings suggest a significant influence of gender, college, and housemaids working hours in the households on respondents' willingness to pay. For example, female respondents with a college degree and households with longer working hours of housemaids are more likely to contribute towards the improvement of the water and sanitation facilities in slums. Though the findings are statistically significant at a 5% level across different estimated models, the regression model exhibits a low goodness of fit.*

* Department of Economics, University of Chittagong, Bangladesh, Email: nurnobi@cu.ac.bd.

**Acknowledgment**

The author is grateful to the Research and Publication Cell of the University of Chittagong, Chittagong, Bangladesh for providing financial support to conduct this research.

# Willingness to Pay to Prevent Water and Sanitation-Related Diseases Suffered by Slum Dwellers and Beneficiary Households: Evidence from Chittagong, Bangladesh

## 1. Introduction

The slum and city are integrated parts of modern life. In Chittagong, 0.36 million people live in the 1814 slums (Daily Star, 2009). The percentage of the slum people is about 18.31 of the total population in Chittagong (New Age Metro, 2005). In the slums, 50% of the population has an energy intake of < 1805 kcal / person/day which means they are living in extreme poverty. Moreover, 11% of the children are affected by diarrhea and malnutrition, and children and women are in high proportion in the slum (BNHS, 2002), which reflects the bad health conditions of the slum people.

These slum people are involved in various important service sectors; such as domestic work, rickshaw pulling, day labor, garment work, cleaning, and many other self-employment sectors (Daily Stars, 2009). As a result, the city dwellers depend on the services of the slum's people to maintain and enjoy the happiness in their lives (Alam, Sultan, and Afrin, 2010. But, very often the slum people get affected by some diseases like gastric pain, dysentery, skin diseases, diarrhea, pain, general weakness yellow fever, etc. because of the water and sanitation-induced problems in the slum (Sabur and Sarkar, 1998). Sometimes they had to embrace death because of these diseases. Lack of safe water is a cause of this whereas water-related diseases are responsible for 24% of all deaths (Water aid /internet). The effect of water and sanitation-induced diseases on slum dwellers can be divided into direct and indirect effects. The direct effect can be measured by adding up their averting activities cost, mitigating activities cost, and loss of income in terms of opportunity cost. But being affected by the disease, the slum people have two indirect effects on society. The first one is the loss of society's production or welfare as the labor supply (by these service people) decreases because of their illness, and the second one is the transmission of diseases to the other people who come into touch with them to enjoy the services.

So, the diseases that domestic workers suffer due to polluted water and unhygienic sanitation negatively affect society by reducing their services and spreading the diseases to others. This is why, to avoid the loss of society's well-being and not to allow the diseases to spread out, society should contribute to improving the water and sanitation facilities in slum areas for the service people (Alam, 2013). In this study, we will try to find the way out (i.e. Willingness to pay) that the city dwellers are willing to pay to improve the water and sanitation facilities in the slum areas to ensure smooth and continuous services for them and to remain in good health.

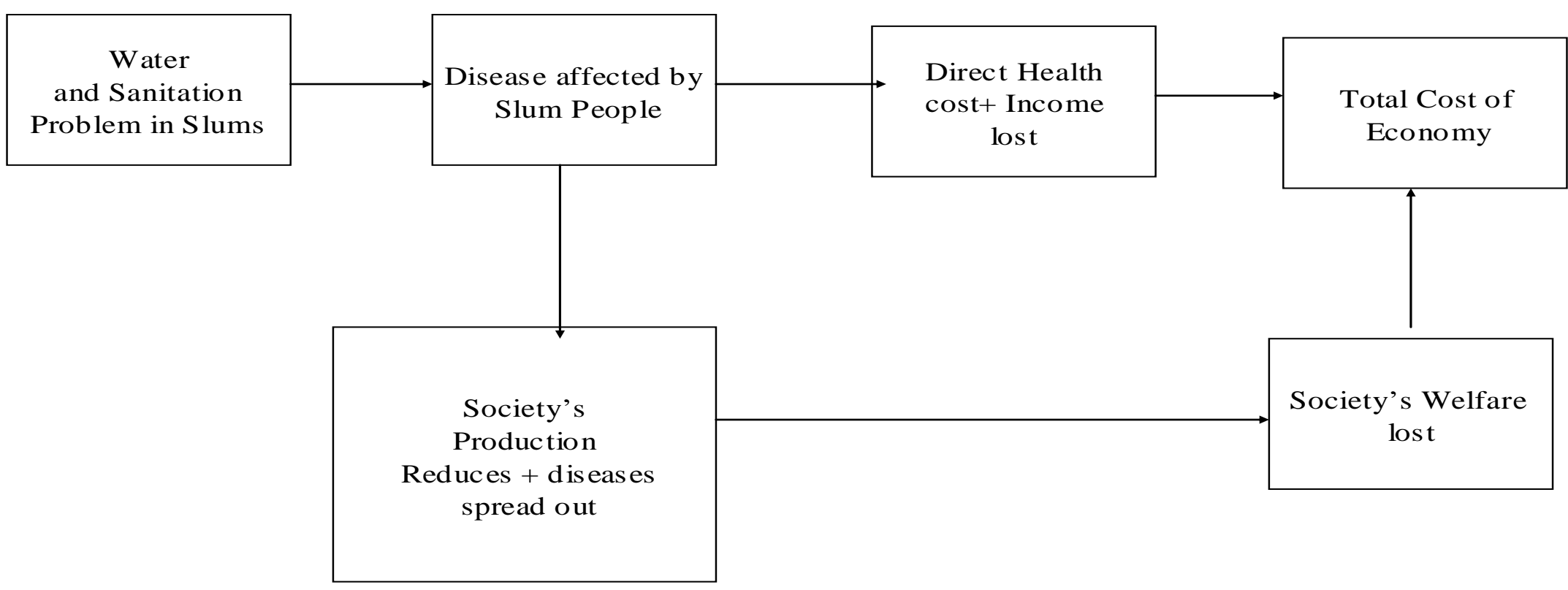

*Figure 1: Conceptual Flow Chart*

## 1.1Rationale of the study

Bangladesh is densely populated by 160 million people in an area of 147,526 square miles where 15.45 M people live in the six metropolitan cities among those 5.44 M populations live in the 9048 slums (Angeles et al, 2005). Like most of the basic needs, safe drinking water and hygienic sanitation are also big problems in Bangladesh. It is seen that 74% of people have access to safe drinking water and 13.5% of rural households use sanitary latrines (IPRSP, 2002). In the urban area, 70% of people live in innumerable slum and squatter settlements (Rahman, 2006). In the slum areas, only 13.5% of people have sanitary coverage while 40% of the total population in the country wash their hands after defecation (DPHP, Dhaka: 2007).

As a result, slum area people often suffer from Diarrhea, Cholera, Dysentery, Dengue, Anopheles, Asthma, Cough, etc. So, slum area people are the victim of environmental hazards which can spread out to others who come in touch with them (Alam, 2013). Because, urban dwellers enjoy the services of slum dwellers, especially women in domestic work and males in rickshaw pulling, day labor, and many other services. An Indian study opined that the slum people are prone to suffer from waterborne diseases like typhoid and cholera, cancer, and HIV/AIDS (Drishti, 2018).

Children in the slums are found to be affected by acute diarrhea because of the lack of proper sanitation facilities and pure and clean water (Hauque et al., 2003). They also found providing pure drinking water can avoid the infection as parents or families are taking care of the children.

Alam et al. (2013) have studied the faculties of the slums in Rajshahi city. They have reported that most of the households (58%) mentioned jaundice as a water-contaminated disease in the slums. The second visible disease in the slums is diarrhea (48%) followed by dysentery (33%) and typhoid (24%). Unsurprisingly, the lack of proper hygiene infrastructure contributes to a higher prevalence of waterborne diseases among slum residents, including jaundice, diarrhea, dysentery, and typhoid (Alam et al., 2013). Even non-communicable diseases (NCDs) are also linked to the lack of sanitation. Therefore, NCDs are prevalent in major parts of Bangladesh, irrespective of different age groups and gender (Roy et al., 2023).

Democracy Watch (2014) has conducted research work among the slum dwellers in Dhaka city, where the findings show that the percentage of people washing their hands after using the toilet and before eating is very low. It also found that most of the slum dwellers are habituated to use hanging latrines. This study has disclosed the fact that most of the slum dwellers suffer from water-borne diseases, like typhoid, diarrhea, and chronic dysentery. This study has identified that the slum people do not understand properly that the use of poor sanitation system and infection of diseases has a causal relationship.

According to the Bangladesh Bureau of Statistics (2015), a considerable number (45.21%) of slum people depend on the water supplied by the authorities which shows that the slum-living people have poor access to clean water, as the piped water is not pure and hygienic in most of the cases. It also reported that 42.19% of slum dwellers are used to have unhygienic toilets. (bdnews24.com, 2015)

A 2009 baseline survey by the Human Development Research Centre shows that the majority of slum people are aware of hygiene rules and regulations but hardly maintain those rules and thus maintaining hygiene is very low. This survey showed that only 1% of respondents said that they wash their hands before eating and 13 % said that wash their hands after wiping a child's bottom. Females in the slums are the main victims of unhygienic lifestyles because of cooking, child care, cleaning the household, and for menstrual (Human Development Research Centre, 2009).

Chittagong, the second largest city in Bangladesh is situated in the southeastern area of the country. The total population of the city is around 4 million where more than 3.6 lakhs (0.36 million) people live in the slum areas. In 1996, the number of slums in Chittagong was 186 where 1, 88,839 people lived in 45,143 households. It is also found that only 5.86% of slum people have sanitary toilets (BBS, 1996). The Daily Star (2017) also published a documentary that stated that a lot of people still live in the slums in Chittagong. According to a survey conducted by Chittagong City Corporation published in Daily Star states that more than two hundred slums are located in Chittagong. These slums are populated by 1 million people living scattered in the city. Unfortunately, the slums are deprived of basic facilities like water, electricity, and sanitation. The report has also included a case from one of the populated slums in Dewanhat. The case stated that there were only four to five toilets for more than two thousand people which depicts the adverse scenario of the slum dwellers in the city (Daily Star, 2017)

*Table 1: Some Characteristics of Slum in Bangladesh*

| Name of the city | No. of Slums | Slum's populations | % of the slum population as city populations |
|---|---|---|---|
| Dhaka | 4966 | 3420521 | 37.4 |
| Chittagong | 1814 | 1465028 | 35.4 |
| Khulna | 520 | 188442 | 19.5 |
| Rajshahi | 641 | 156793 | 32.5 |
| Barisal | 351 | 109705 | 30.1 |
| Sylhet | 756 | 97675 | 27.1 |

Source: *Angeles et al., 2009 (International Journal of Health Geographic)*

In contrast, in 2005 the number of slums reached 1814 where 18.31 % of the total population of the city lived in the slum areas (New Age Metro/Internet). According to Angeles et al. (2009), the number population in Chittagong city is 4,133,014 where 1,465,028 people live in 1814 slums (Table 1).

Though the people living in slums have poor access to basic services (i.e. pure drinking water and hygienic-sanitary), they are usually involved with some important service sectors which are essential for the city dwellers (Alam, 2013; Alam and Hossain, 2018). About 79% of the working people in the slums are involved in informal service sectors in 15 different professions (Chowdhury, 1989). If the slum dwellers suffer from water and sanitation-induced illness, it will ultimately affect the city dweller's utility through the services they receive from the slum people. Almost 19% of slum people are involved in rickshaw pulling and about 23% are involved in kitchens in hotels & restaurants and in domestic work (Alam, Sultan, and Afrin, 2010; Chowdhury, 1989). Therefore, because of the dependency of the city dwellers on the slum people, this study aims to explore whether the city dwellers are willing to pay to improve the water and sanitation problems in the slums for the betterment of their own utility function.

The findings of the study can be used to provide guidelines for policymakers, non-governmental organizations (NGOs), and development partners to take proper initiatives. It can help them to undertake water and sanitation projects in slum areas considering the economic importance of these problems.

Moreover, the Delhi Summit on Water and sanitation, 2008 has declared the inevitability of ensuring safe water and sanitation for 1000 million people of the eight South Asian countries by 2012. Being a founder member of the South Asian Regional Cooperation (SARC), Bangladesh has given priority to safe water and sanitation facilities. Therefore, the findings of this study have huge policy implications that could be helpful in framing the targeted policy of ensuring water and sanitation facilities in the country.

### 1.2 Hypothesis /Research Question:

The hypothesis of this study is, that access to safe water and hygienic sanitation decreases disease for slum people which will lead to a social net gain (reducing negative externality) for the households who receive the services of the slum people.

## 1.3 Aims and Objective of the Research

The broader objective of this study is, how much the urban people (Society) are willing to pay to reduce the water and sanitation problems in slum areas which makes negative externality to them. However, the specific objectives of the study are as follows:

i) To analyze the negative externality of water and sanitation-induced health problems of slum people to the society in terms of welfare loss.

ii) To estimate the impacts of the factors that motivate Willingness to Pay (WTP) of the households who are dependent on the services of the slum people; and

ii) To suggest policy guidelines and recommendations for improving the water & sanitation-induced health problems in the slum areas.

## 1.4 Organization of the Study

The remaining sections of the study have been organized as follows; section 2 presents the literature review. Section 3 presents the methodology and Methods. Section 4 of this study analyzes the results and discussions. Section 5 is the concluding remarks and limitations of this study and finally, the references have been articulated.

## 2. Literature Review:

There are many studies that broadly focus on the issue under consideration. The most relevant among them are reviewed below:

The lack of pure drinking water and sanitation problems in the slums in Bangladesh is well known to everyone and is supported by various studies (Alam, 2013; Alam and Hossain 2018). Ruma (2009) studied that South Asia (especially Bangladesh) is in sanitation problems. In this region, 1000 million people live in high risk about 0.50 million children's death is caused by diarrhea, and 250 million people are affected by water and unhealthy sanitation-based diseases.

In addition to sanitation-based diseases, the slum people in Bangladesh also suffer from environmental hazards which may spread out and may cause serious health damage to other people. Alam (2013) concluded that slum area people are suffering from environmental hazards that may be spread out to the other people who come in touch with them as it is almost clear that beneficiary people of the urban area are enjoying happiness by dint of the services of the slum area people and these diseases and externalities of the slum people have to be suffered by the urban beneficiary people.

Uddin, et.al (2006) found that people in the slum area suffer from the lack of health care more than the other people in Bangladesh. In the slums about 70% of mothers suffer from malnutrition and anemia and less than 40% of the population has access to basic health care.

Mostafa, et al (2006) showed that the situation of women in the slum area is worse than that of mainstream women where only 24% of working women have access to safe drinking water. Those who use pond water do not boil water and most of them use traditional unhygienic latrines though 30% of women use proper hygienic latrines. Thus, they suffer from Diarrhea and gastric Ulcers. Anemia, Cough, etc.

Alam (2005) aimed to work on the situation of water and sanitation in the slums of Dhaka, Bangladesh. In the study, he intends to find out the parent's (slum Households) willingness to pay to avert diarrhea attacks on their children originating from polluted drinking water and unhygienic sanitation. However, the willingness to pay of the slum people is very low, and therefore problems associated with water and sanitation are deteriorating. Alam (2013) and Islam (1994) found a similar result in their study and concluded that the situation of water sanitation, and environmental concerns in slums of Dhaka is worsening due to low willingness to pay.

Chowdhury (1989) found that slum people were 208000 in 1989 in the Chittagong city area where 79% of all working people were self-employed and mostly involved in the informal service sector in 15 different professions. 19% of them are involved in rickshaw pulling and almost 23% are involved in kitchens in hotels, restaurants, and domestic works. Among them 32% of working people are female. He concluded that 19 people use one latrine and use polluted water for washing and bathing. As a result, Diarrhea and Malaria are common diseases in slum areas.

National Surveillance Project (NSP) Bulletin, No.9 (2002) emphasized better food security health services, and sanitation facilities for urban poor people in Bangladesh. Both for humanitarian and economic reasons it involved the Government of Bangladesh, NGOs, and donor agencies.

Nam and Son (2004) studied household demand for improved water services in Ho Chi Minh City. They compared contingent valuation and choice modeling estimate. Dwight, et al. (2005) estimated the economic burden from illness associated with coastal water pollution in the recreation areas of California, U.S.A

The reviewed studies show that no specific study on water and sanitation-induced diseases and their external effects has been conducted at the national level and also at the regional level in Bangladesh. Therefore, a significant study gap remains in this area. With this consideration, this study endeavors to explore the water and sanitation-induced diseases in the slums and their indirect effects on the city dwellers in terms of their willingness to pay (WTP).

## 3. Methods and Methodology

### 3.1 Data

The study is based on primary data. There are two types of data that have been collected from two groups of people. First, the housemaids who work in various households are considered the first group of respondents of this study. The data from this group has been collected through Key Informant Interviews (KIIs). A total of 10 KIIs were collected from two densely populated slums in Chittagong. The first one is located at Bogarbil (6 KIIs) and the second one is located at Sholoshahar (4 KIIs). With well-structured KII guidelines, the respondents were asked several questions on their socio-economic background, the condition of water and sanitation facilities in the slums, its impact on their health, and their willingness to pay for the provision of proper water and sanitation facilities.

The second group of respondents is the households where slum women work as housemaids. The primary data has been collected from households in the city area. Houses where slum women work are considered as the hub of the external effect of the housemaids. Various areas of the Chittagong Metropolitan have been considered as the study areas of this study. The targeted population of this study is the household who depend on the housemaid for their services. The initial plan of the data collection was to select 10 dwellers from each of the targeted slums including males and females to assess the health problems induced by water and sanitation problems. Then the households or the working places where the selected persons work will be systematically selected as the sample unit. In the case of the selection of an area as a slum, the city corporation's directory has been considered to avoid ambiguity. From both of the sample units, personalized interviews were conducted to collect the required information. But, after the spread of the coronavirus pandemic, most of the services were shut down and in-person interviews based on the selection of the occupation were impossible. Therefore, the study focused on a specific occupation; the services of the housemaid. Finally, sample data has been collected from households that employ women of the slums for assisting with their households' chores. The data collection was based on a well-structured questionnaire interview. A group of data enumerators with knowledge of research methodology and proper training in interviewing the respondents have been employed for two weeks. The details of the primary survey data are presented in Table 2.

*Table 2: Area-wise sample distribution*

| Serial | Area | Frequency | % |
|---|---|---|---|
| 1 | Agrabad Neighborhood | 46 | 17.76 |
| 2 | Andarkilla Neighborhood | 26 | 10.07 |
| 3 | Bahaddarhat-Chandgaon Neighborhood | 20 | 7.73 |
| 4 | Chawkbazar- Bakalia Neighborhood | 29 | 11.19 |
| 5 | Bibirhat Neighborhood | 21 | 8.11 |
| 6 | Halishahar Neighborhood | 29 | 11.20 |
| 7 | Khulshi Neighborhood | 22 | 8.49 |
| 8 | Lalkhan Bazar Neighborhood | 22 | 7.34 |
| 9 | Nasirabad Neighborhood | 26 | 9.65 |
| 10 | New Market and Kotowali Neighborhood | 19 | 6.95 |
| | Total | 260 | 100 |

The enumerators collected the data using a face-to-face interview technique. For each of the data GPS coordinates have been collected which it possible to see how evenly data was collected from the city (Figure 2). These coordinates also explain that the data is representative of the city dwellers as the spread of the data is wider.

They also observed the respondents' attitudes as a means of a participatory approach. Since the data was collected in February 2021, the enumerators maintained the proper guidelines of social distancing, and wearing masks and interviewed the respondents staying outside of the houses/flats. The initial sample size was 200, but with careful observation of the various stratum, the final sample stands for 260. The details of the data collection are shown in Table 2 above.

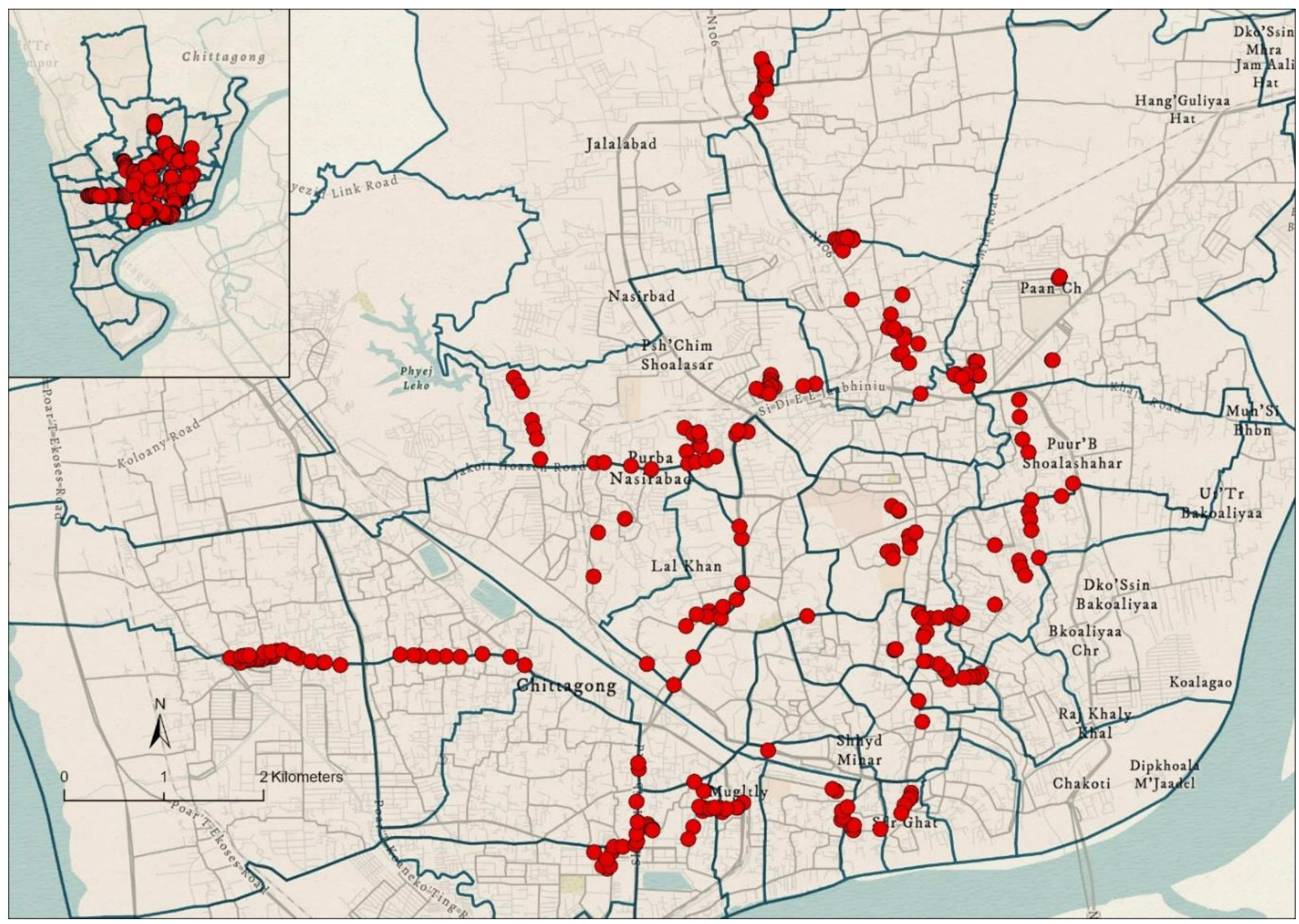

*Figure 2: Study Area Map*

## 3.2 Econometric Model

This study will use the Contingent Valuation Method (CVM) to investigate whether a household is willing to pay (WTP) to avoid water and sanitation-induced diseases or not. In this regard, what factors motivate households to show their willingness to pay will be identified. In addition, if they want to pay, how much they will be willing to pay will also be estimated. The WTP tool is a widely used indirect stated preference tool in economics. With the help of the WPT of the city dwellers, this study will state the probable contribution to improving the water and sanitation facilities in the slums as a social responsibility to improve the quality of lives of the slum people which will in turn produce benefits for households. The cost of poor water and sanitation facilities in the slums that affect people's health function or utility functions are estimated through the damages people suffer both directly and indirectly. The direct and indirect health effects of water and sanitation-induced diseases are estimated through the following ways:

### 3.2.1 Direct effect:

The direct effects of water and sanitation-induced problems in slum areas can be measured by the Willingness to Pay (WTP) of the slum people. The direct effect of the water and sanitation-induced problems is borne by the slum households and therefore increases their sick days. If a sick day arises, it reduces their earnings which is their direct loss. Therefore, the slum people were asked about their willingness to pay. In the WTP question, they were asked 'how much they would pay not to be sick or to avoid the probability of being sick'. The underlying methodology of the direct effect of water and sanitation-induced diseases can be understood by following health production function:

The health production function of the working people (slum people) is;

$$S = S(D, A, M, H) ------------------------------(1)$$

Here, S=Sick days, D =Disease, A=Averting Activity, M=Mitigating Activity, H=Household Characteristics. In addition, the disease function could be expressed as,

Disease Function; D = (WL, SL, A, Aw) --------------------------------------- (2)

WL= Lack of access to safe water, SL = Lack of Proper sanitation facilities, awareness.

These equations determine the utility function of the slum households. Thus, maximizing the slum household's utility function is an objective function that they will try to maximize using the constraint of money income;

Utility Function: U= (X, L , S ; H)------------------------------- (3)

Subject to: I+ Pw (T-L-S) = X + Pa. A + Pm. M --------------- (4)

Here $P_w$ refers price/wage of the working time, T refers to total time (Month/day), and L refers to leisure. Now, by solving these equation systems, we can find the values of A. M and S which can help to assess the direct cost of water and sanitation-induced diseases. The direct cost is thus the maximum willingness to pay of the households to avoid the diseases that reduce their utility in terms of monetary value, such as,

WTP= Pw (ds /dD) +Pa (δA*/ δD) + Pm ((δP*/ δD)- (δU/ δS)/λ. dS /dD ------- (5)

And solving the equation we find the values of A, M, and S which can help us to assess the direct cost of illness of the slum people.

### 3.2.2 Indirect effects

Society is the ultimate sufferer of the water and sanitation-induced diseases in the slums in the city. As the city dwellers depend on the slum people for various services, they are sufferers through the services. Therefore, to avert the indirect effects or the external effects of the diseases that slum dwellers suffer can be measured in two ways.

First of all, the effect which is marketable, that is a reduction in society's production or utility due to the shortage of labor supply from the slum dwellers can be measured by the Surplus Method. That is, to avoid the external effect the society will pay the amount equivalent to what they lose due to the reduction of the labor from the slum dwellers.

Secondly, the effect is not marketable, that is if the presence of sick workers (service providers, especially domestic workers) affects the household. If any member of the household gets sick (service receiver) they must spend money to get a cure. Therefore, how much the service receiver is prepared to pay in order to prevent sickness (in the presence of the slum people) will be estimated by the Willingness to Pay (WTP) approach (Alam, 2013; Alam and Hossain 2018). That is to maintain their utility or welfare; the society will pay to improve the water and sanitary facilities in the slum areas. Here WTP of the households/service receivers depends on their ages, Sex (male or female; dichotomous variable), financial condition of the respondent, level of education, occupation of the respondents, and types of workers (full-time/part-time), etc. The empirical model is assumed to follow the logistic model. A logistic is defined as follows:

$$L_i = \ln\left(\frac{P_i}{1-P_i}\right) = \beta_1 + \beta_2 X_i + u_i \text{------------------------------} (6)$$

Here, Li stands for log of odd ratios, Pi for probability, Xi is a vector of the explanatory variables and ui is an error term. The Li (log of odd ratio) is linear in parameters which could be estimated as,

$$P(Y|x_1 \ldots\ldots x_k) = f(x_1 \ldots\ldots x_k) \text{------------------------------} (7)$$

Here Y is a binary variable (Equation 7) and Y=1 stands for households' willingness to pay for avoidance of external effects of the slum people and Y=0 refers otherwise. Here, $f$ refers logistic distribution function (Science Direct, 2021). This logistic distribution function transforms the regression into the interval of 0 and 1. The logit model can be further defined as,

$$\log it \quad (P(Y=1|x_1 \ldots\ldots\ldots x_k)) = \beta_0 + \beta_1 x_1 + \beta_2 x_2 \text{---------} + \beta_k x_k) \text{----------} (8)$$

Now, following equation 8 above, the empirical model of this study stands as follows;

$$\log it(P(WTP = 1 | x_1 ..........x_k))$$
$$= \beta_0 + \beta_1 Sex_1 + \beta_2 Education + \beta_3 FamilyIncome + \beta_4 Occupation + \beta_5 TypesofWorkers) ------(9)$$

Here WTP refers willingness to pay, Sex is a dummy variable (1= female and 0= otherwise). Education refers the years of schooling which is also a binary variable. The value of education equal to 1 denotes that the respondents have 12 years and above schooling. Family income is a quantitative variable that refers to the monthly total family income of the respondents. Occupation is another dummy variable whereas its value equal to 1 refers to the respondent being a housewife and zero for others. Types of workers are used to identify whether the worker is a full-time (Types of workers = 1) worker or part-time worker (Types of workers = 0).

## 4. Results and Discussion

The result of this study has been summarized in two sections; descriptive and quantitative. In the descriptive section summary of the findings of the two respondent groups has been presented. In the first sub-section of the descriptive finding, the Key Informant Interviews were summarized with the qualitative matrix. In the second sub-section of the descriptive analysis, the findings of the survey of the households (city dwellers) have been summarized as descriptive statistics. Finally, the empirical findings of the study have been summarized in the quantitative section.

### 4.1 Findings of the Key Informant Interviews

As mentioned earlier, there were 10 KIIs. The KIIs were chosen from two slums in Chittagong. The respondents of the KIIs are the group of the population who suffer from water and sanitation-induced diseases in the slums. These respondents are providing services as housemaids to the city dwellers. Almost all respondents were married except one. They are working at households to support their families. The major occupations of the husbands of the respondents were rickshaw pulling, day laborers, and small traders. In the case of the respondents living premises (slums), all women are working either in garment factories or in households as housemaids to support their families. The reasons that the respondents mention for doing a housemaid job are the death of the husband, sickness of the husband, and poverty.

*Table 3: Major findings of KIIs*

| KIIs | Cleaning the slums | Quality of water | Facilities of drinking water | Toilet facilities | Common diseases | Want improvement | Who will pay |
|---|---|---|---|---|---|---|---|
| KI-1 | Cleans regularly | good water | Good drinking water | 1 bathroom for 2 families | Fever and cold | Yes | government |
| KI-2 | Cleans but not regularly. It's stench. | deep tube-well, far better | Deep tube-well water | Wash hands | fever, joundis | No | landowner |
| KI-3 | Cleans regularly | Good | same water | 2 bathrooms for 30 individuals | Fever and cold | No | self but 10 taka/month |
| KI-4 | Cleans but stench. Don't clean regularly | No good water | same water | 3 bathrooms for 8 families/ 25 persons | fever, jondis | Yes | government |
| KI-5 | Cleans one after another day | motor water | collect when a motor starts | 3 bathrooms for 25 families | fever and cold | Yes | Self but 10 taka/month |
| KI-6 | Cleans but not regularly. It remains stench. | Good water and stays all time | same water | yes, 8 families, 45 person use 2 toilets | Fever and cold | Yes | Self and 50-100/month |
| KI-7 | Cleans but not regularly. It remains stench. | motor water, very red with high iron | iron water boil and drink | Yes | Fever | Yes | Don't want to pay |
| KI-8 | Cleans regularly | Good | clean water | Yes | fever and cold | Yes | Don't want to pay (poverty) |
| KI-9 | Cleans sometimes | Not good | polluted water | 1 for 30 families | fever and cold | Yes | Self but 10 taka/month |
| KI-10 | Cleans sometimes but stench. | deep tube-well water and stay long | clean with alum | 3 family use 1 toilet, no water facilities | fever and cold | Yes | Don't want to pay (poverty) |

Therefore, they are relying on their income from this service to support their food costs, accommodation and others relevant costs as an assistance to their husband to maintain their life and livelihood. About 20% of respondents were found to be involved as housemaids after the pandemic spread out. 90% of the respondents replied that they have other family members who do the same job. Almost all respondents were found to work as part-time and their average income was 4,000 taka. The major findings of the KIIs have been summarized through the qualitative matrix in Table 3 above.

From Table 3, it is seen that slums are cleaned but not on a regular basis, and thus stench remains there. In terms of water supply, the respondents replied that they have good water supply and it is mostly from deep tube wells. In some areas iron is mixed with the water and respondents need to take extra measures like cleaning with alum. The toilet facilities in the slums are very poor. 8-25 families' 2-3 toilets which are cleaned weekly. Moreover, there is no water facilities in the toilets. The most common diseases in the slums were reported as fever, cold, and jaundice. The respondents want the improvement of the water and sanitation facilities in the slums but they don't want to pay for it. Only 2-3 out of 10 respondents showed their willingness to pay but a very negligible amount. The main argument they have is their poverty. Secondly, they demand it to be provided by the government. The rationale for the respondents not showing their willingness to pay for the improvement is not having financial loss as salary cut by the households where they work. Therefore, being sick has no opportunity cost to the respondents.

## 4.2 Descriptive Statistics

Descriptive statistics shows the reliability and representativeness of the data with its population. Various properties of the data have been depicted in this section to see how representative and reliable the sample data is with the population.

### 4.2.1 Age of the Respondents

Table 4 below shows the age distribution of the respondents. As shown in the table the middle age group has the highest representation in the sample. Age group 26-30 to 46-50 has about 72% representation in the sample which shows that experienced and elderly population of the society have been included in the sampling. These groups of people can make proper decisions based on their experiences and observations.

*Table 4: Age distribution of the respondents*

| Age group | Frequency | % |
|---|---|---|
| 20-25 | 19 | 7.57 |
| 26-30 | 40 | 15.95 |
| 31-35 | 39 | 15.29 |
| 36-40 | 40 | 15.94 |
| 41-45 | 40 | 15.94 |
| 46-50 | 28 | 10.76 |
| 51-55 | 27 | 10.38 |
| 56-66 | 18 | 6.92 |

### 4.2.2 Gender

Gender is an important characteristics of the society. Gender distribution (Table 4) of this study shows that about 85% of respondents were female whereas 15% of respondents were male. The gender distribution shows that it is discriminatory as women's participation is dominating.

*Table 5: Gender distribution*

| Sex | No. | % |
|---|---|---|
| Female | 222 | 85.38 |
| Male | 38 | 14.62 |
| Total | 260 | 100% |

But it is not unlikely since women are maintaining the household chores and therefore they are the best observers of the research problem. Thus, the gender participation is representative of society.

### 4.2.3 Educational Qualification

The educational qualification of the respondents has been summarized in Table 5 below. Out of six categories of education, most of the respondents were found to have an undergraduate degree which is about 27.31%. After the graduate HSC and SSC got the highest level of education, respectively.

*Table 6: Educational Background*

| Level of education | Frequency | % |
|---|---|---|
| Below SSC | 9 | 4% |
| SSC | 43 | 19.54% |
| HSC | 56 | 25.45 |
| Graduate | 71 | 27.31 |
| Doctor | 3 | 1.36% |
| Master | 38 | 17.27% |
| Total | 220 | 100% |

### 4.2.4 No of Children

It has been found that households have maximum 6 children with mean value of 3 children (2.75) which is likely as the fertility rate in Bangladesh is 2.4 children per woman (worldometers, 2020). However, most of the households are having two or three children (Table 7). This table also depicts that every household has school-going children with a mean value of 2.

*Table 7: Number of Children and schooling per household*

| | Children per household | | School Going Children | |
|---|---|---|---|---|
| Number of Children | Frequency | % | Frequency | % |
| 0 | 2 | 0.84 | 14 | 6.25 |
| 1 | 23 | 10.46 | 59 | 26.34 |
| 2 | 85 | 35.56 | 91 | 40.63 |
| 3 | 72 | 30.13 | 44 | 19.64 |
| 4 | 41 | 17.15 | 14 | 6.25 |
| 5 | 10 | 4.18 | 2 | 0.89 |
| 6 | 6 | 2.51 | ------ | ------- |
| Total | 239 | 100 | | |

### 4.2.5 Occupation of the respondents

In terms of occupation, the findings show that about a major portion of the respondents are housewives. Job holders counted for 18.70% followed by the business. About 16.67 % of the respondents are involved in others including various professionals like Bankers, doctors, advocates, and engineers (Table 8).

*Table 8: Occupation of the respondents*

| Occupation | Frequency | % |
|---|---|---|
| House wife | 104 | 42.28 |
| Job holders | 46 | 18.70 |
| Business | 39 | 15.85 |
| Teacher | 16 | 6.50 |
| Others | 41 | 16.67 |
| Total | 246 | 100 |

### 4.2.6 Monthly Family Income of the Respondents

When it comes to the income of the respondents, most of the respondents replied with no income. It is mentionable that Table 7 above shows that 42% of respondents are housewives who definitely believe that they have no direct income though they are contributing to the service sector through their family. But when it comes to their family income, 255 out of 260 respondents have replied to this question. Among these respondents, about 42% have 20000-

50000 taka as total monthly family income, about 26% have 50000-70000 taka, 27% have 70000-100000 taka and only 7% have above 100000 taka monthly family income (Table 9).

*Table 9: Monthly income of the respondents*

| Income group | Frequency | % |
|---|---|---|
| 20000-50000 | 108 | 42.35 |
| 50000-70000 | 60 | 23.53 |
| 70000-100000 | 70 | 27.45 |
| 100000+ | 17 | 6.77 |
| Total | 255 | 100 |

However, the mean income of the respondents is about 71620 taka which is more than the average monthly income (26,000 taka) in the country (salaryexplorer, 2021).

### 4.2.7 Number of Family members

The average size of the family is 5. The family size also matches the number of children as most of the families have 2 or 3 children (Table 7). The minimum members of the family is 2 whereas the maximum number is 23. There is only one frequency for the maximum family and the frequency of having more than 7 family members is 18 only. It resembles that the number of joint families is very low (Table 10) below.

*Table 10: Size of the families*

| Number of family members | Frequency | % |
|---|---|---|
| 4 | 98 | 40.16 |
| 5 | 77 | 31.56 |
| 6 | 35 | 14.36 |
| 7 | 16 | 6.56 |
| More than 7 | 18 | 7.35 |
| Total | 244 | 100 |

### 4.2.8 Working condition of the Housemaids

The data shows that out of about 99% of households in Chittagong City have one housemaid and only two households have two housemaid. Out of these families, 8% of families have employed a full-time house maid and almost 92% of them depend on a part-time housemaid (Table 11). The table describes that about 100% of households are somehow dependent on the services of the housemaid.

*Table 11: Working statistics of the housemaid*

| Type of housemaid | Frequency | % |
|---|---|---|
| Full time | 22 | 8.46 |
| Part time | 238 | 91.54 |
| Total | 260 | 100 |

The full-time housemaids are employed for 24 hours as they are staying with the households. The data shows that full-time workers are working a minimum of 6 hours to a maximum of 24 hours (they stay at the employers' houses), according to the respondents. However, the average working hours for full-time housemaids are found to be 22 hours though the maximum frequency (20) for working hours of the housemaid is 24.

*Table 12: Wage distribution of the Full time workers*

| Monthly wages | Frequency | % |
|---|---|---|
| Below 3000 | 4 | 18.18 |
| 3000-4000 | 10 | 45.45 |
| 5000-7000 | 6 | 27.27 |
| Overall cost | 2 | 9.09 |
| Total | 22 | 100 |

According to the respondents, most of the full time house maid get a monthly wages of 3000-4000 taka. It also includes their three times food and clothes. About a quarter of the housemaid receive the highest wages of 5000- 7000 taka only (table 12).

In contrast, the part-time housemaid works from 30 minutes (minimum) to 12 hours (maximum) with an average of 2.39 hours in a day in the households. Table 13 shows that most of the part-time house maid work for 2-3 hours daily (56.17%).

*Table 13: Working hours of the part time house maid*

| Working hours | Frequency | % |
|---|---|---|
| Less than 2 hours | 69 | 29.36 |
| 2-3 hours | 132 | 56.17 |
| 4-5 hours | 22 | 9.36 |
| 6-12 hours | 14 | 5.95 |
| Total | 235 | 100 |

The minimum wages of the part time house maid is reported as of taka 500 whereas the maximum wage is 6000 taka with an average of 2067 taka. Table 14 below shows the distribution of the wages of the part-time housemaid. It also shows that most of the housemaids (48.5%) receive below 2000 taka followed by 2000-3000 taka (41.2%).

*Table 14: Wages of the part-time house maid*

| Wages | Frequency | % |
|---|---|---|
| Below 2000 | 113 | 48.49% |
| 2000-3000 | 96 | 41.20% |
| 3000+ to 6000 | 23 | 9.87% |
| Total | 233 | 100% |

The respondents also replied that most of the households employ the house maid for cleaning the houses and washing the clothes (50.38%). On top of this 20% respondents replied that they require the housemaid including for manage all household chores. Below 10% of respondents replied that they depend on the housemaid for cooking. Only 4% of respondents rely on the housemaid to take care of children. According to the respondents (88%) the housemaids are staying outside of the residence that is they are providing these services from their own living arrangements. About 20% of households reported that their housemaid stayed in a rented house but they did not mention any specific place. However, 14% of households reported that their housemaid lives in the Bogarbil; which is reported as the largest slum in the city. About 106 respondents replied to the question on their knowledge about the living conditions of the housemaids.

*Table 15: Working tenure of the current housemaids*

| Tenure | Frequency | % |
|---|---|---|
| Less than 6 months | 99 | 38.08 |
| Less than 1 year | 46 | 17.69 |
| 1-3 years | 78 | 30.00 |
| More than 5 years | 22 | 8.46 |
| More than 10 years | 15 | 5.77 |
| Total | 260 | 100 |

And those who replied, 49% opined that they know about the living condition of the housemaids. It states that less than 25% of total respondents know about the living conditions of their housemaid only. About 38% of respondents replied that their current housemaid is working for less than 6 months and less than one year is 18%. However, a good number (30%) of respondents replied that the current housemaid is working for 1 – 3 years (Table 15).

### 4.2.9. Ownership of houses

The majority of the respondents (about 66%) are living in a rented house whereas one-third of them are living in their own house/ apartment (Table 16). A few portion of the respondents were found to be stayed in government quarters.

*Table 16: Ownership of houses*

| Housing types | Frequency | % |
|---|---|---|
| Rented | 170 | 65.89 |
| Own | 83 | 32.17 |
| Government apartment | 5 | 1.94 |
| Total | 258 | 100 |

For those who live in rented houses, the average house rent per month they pay is about 15300 taka. Though the average monthly house rent is about 15000 taka it ranges between 7000 to 30000 taka.

### 4.2.10. Households sickness and willingness to pay

When respondents were asked to mention whether any member of their family suffered from any infectious diseases after the current housemaids were employed, about 98% of respondents replied that none of their family members became sick after the current housemaid was employed. Diseases they suffered were coronavirus infection and diarrhea. In response to a question on knowledge of housemaids on water and sanitation, about 73% of respondents replied that they believe that the housemaid has enough knowledge on water and sanitation. Almost all respondents have replied that the housemaid drinks pure drinking water, boiled water, or WASA-supplied water. They also opined that the housemaid washed their hand with soap and their housing premises were clean. Surprisingly, there came a controversial response from the respondents which is they had taken any measures if the housemaid had no knowledge about water and sanitation, and it shows 29% of respondents replied that they had taken any measures.

*Table 17: Willingness to Pay Questions*

| Items | Yes (%) | No (%) | Total |
|---|---|---|---|
| Suffered from diseases (%) | 1.98 | 98.02 | 100 |
| Aware of the knowledge of housemaid | 72.97 | 27.03 | 100 |
| Have taken any measures | 29.11 | 70.89 | 100 |
| Know that housemaids knowledge on water and sanitation may affect them | 89.53 | 10.47 | 100 |
| With this knowledge, have taken measures | 53.91 | 46.09 | 100 |
| Will support the government if any measure is taken to provide water and sanitation facilities in the slums | 98.46 | 1.64 | 100 |
| Willingness to pay fees | 56.86 | 43.14 | 100 |

The measures they have taken were reported as awarded her to be clean, drink boiled or pure water, wash hands before entering the houses, maintain sanitation, etc. Among the respondents, about 90% know that housemaids' knowledge about water and sanitation can affect the households, and thus to make sure that knowledge about 54% of the respondents have taken further measures. The actions they have taken were also awareness building on using proper sanitation, drinking pure water, and practicing cleanliness. In this regard, most of the respondents (98.65) have shown their willingness to support the government if the government takes any measures to provide pure drinking water and sanitation facilities in the slums and living arrangements of the workers. In support of voting the government on the previous facilities, about 57% of respondents have agreed to pay fees to the government or the city corporation if they take any measures. Those who wanted to pay fees to the government or the city corporation, were divided into paying the fees in monthly, annually, and as a lump sum (Table 18).

*Table 18: Payment Vehicles and amount*

| Way of payment | Frequency | % | Mean amount (Tk.) |
|---|---|---|---|
| Monthly | 50 | 39.37 | 133.67 |
| Yearly | 39 | 30.71 | 414.86 |
| Lump sum | 36 | 28.35 | 448.55 |
| Others | 2 | 1.58 | ------- |
| Total | 127 | 100 | |

About 39.37 of the respondents agreed to pay by month while 31.71% expressed their views to pay annually, and 28.38% of respondents wished to pay a lump sum amount. The mean fees that the respondents agreed to pay are about 134 Taka per month, 414.86 Taka per annum, and 448.55 Taka as a lump sum. However, the average payment that the respondents are willing to pay is about 412 Taka irrespective of their payment vehicles.

In contrast, some respondents haven't shown their willingness to pay as they believe that it is the responsibility of the government (41.07%), the responsibility of the city corporation (8%), and the responsibility of both the government and City Corporation (50.86%).

The respondents were also asked to mention their monthly medical costs to the family and it shows that on average per household's medical/ treatment costs move around 5000 Taka. Surprisingly the standard deviation of the mean value of medical costs is very high (15911) as the individual family costs vary most frequently. This cost ranges from 200 Taka to 25,000 Taka per month.

After asking about the per-household medical/treatment costs, with another hypothetical scenario, they were asked whether they would pay the fees/charges if they realized that the implementation of a water and sanitation project in the slums reduces their medical costs.

*Table 19: Households' willingness to pay after any project is taken*

| | Households willingness to pay | | | | Total |
|---|---|---|---|---|---|
| Vehicles | Yes (%) | Average amount (if yes) | Didn't mention the amount | No | |
| Monthly | 16.47 | 162 Taka | | ---- | |
| Yearly | 17.80 | 332 Taka | | ---- | |
| Lump sum | 39.62 | 337 Taka | | ---- | |
| | 73.90 | | 4.41% | 26.10 | 100 |

In response to this question, 73.90% of respondents showed their willingness to pay for the water and sanitation project in the slums.

But, when they were asked how they want to contribute the fees/charges, only 69.46% respondents replied this question. Table 19 shows that 39.62 % of the respondents wanted to contribute as a lump sum amount followed by a yearly payment (17. 80%). On an average respondents wanted to pay 337 taka as a lump sum amount to improve the water and sanitation facilities in the slums. Of course, 100% of the respondents replied that they believe that raising social awareness in this regard is essential.

## 4.3 Empirical Estimation

In this section, the estimated results have been summarized based on equation 9 in section 3.2.2. The estimated results are shown in Table 20 below. To have the most precise estimation, 4 different models have been estimated and in all the estimations sex and education are found to be significant below a 10% level of significance. All other variables are statistically insignificant. Though the coefficients are statistically insignificant, signs of the coefficients of the variables are theoretically consistent. For example, the coefficient of sex is 0.67 (in model 2) which is also statistically significant at the 5% level which means that the female respondents are 0.67 times more likely to pay for the water and sanitation project (if it is taken at all) in the slums. Similarly, the coefficient of education 0.388 (in model 1) denotes that respondents who have 12 years or more schooling are more likely to pay (have willingness to pay) for the water and sanitation facilities in the slums for the slum people who provide services to the households. However, the coefficient of education is not statistically significant below 10%. The estimations show that the p-value of most of the coefficients of the explanatory variables is close to 0.2.

*Table 20: Estimated Results*

| Coefficients | Model 1 | Model 2 | Model 3 | Model 4 |
|---|---|---|---|---|
| Constant | -0.790**<br>(-1.76) | -0.755<br>( -1.69) | -0.678<br>(-1.50) | -1.918**<br>(-1.97) |
| Sex | 0.547<br>(1.41) | 0.671*<br>(1.83) | 0.66*<br>(1.79) | 0.66*<br>(1.79) |
| Education | 0.388<br>(1.43) | 0.340<br>(1.28 ) | 0.333<br>(1.25) | 0.333<br>(1.25) |
| Family income | 0.000003<br>(1.09) | 0.000003<br>(1.11) | 0.000003<br>(1.07) | 0.000003<br>(1.07) |
| Occupation | 0.283<br>(0.99) | - | - | - |
| Full day/part time | - | - | -0.62<br>(-1.35) | 0.62<br>(1.35) |
| Prob > chi2 | 0.132 | 0.132 | 0.112 | 0.112 |

The chi2 value is also close to be significant about 10% level, especially model 3 and model 4. Therefore, the estimated results could be accepted as they can provide some idea about the households' willingness to contribute to the development of the living conditions of the slum people in an effort to avoid external health effects.

## 5. Conclusion and limitations

Supply of pure drinking water, usable water, and proper sanitation facilities are not well provided in the slums in Bangladesh. As a result, due to this limitation, the slum people may be infected by the diseases associated with water and sanitation problems. Because of the dependency on the slum people, the city dwellers are at risk of contamination by these diseases. Therefore, this study explores the willingness to pay for households who depend on the slum area's services. The respondents, on average, have agreed that water and sanitation facilities in the slums are deplorable, and therefore the housemaids who provide services to the households may not maintain hygiene. They may also be carriers of various diseases. Most of the respondents believe it, and thus many of them (56.86%) have shown their willingness to pay. Respondents who disagreed on willingness to pay claimed it was paid by the government or local government authority like City Corporation. A similar opinion was also discovered from the KIIs with the slum people, who demanded that the government pay for these services. They argue that they hardly manage their livelihood and thus could not pay additional fees to the government.

Most interestingly, most of the KIIs opined that they do not suffer very much from water and sanitation-induced diseases, and therefore, they do not have much concern. However, this study has several limitations; firstly, a similar empirical estimation (like Table 17) was projected to assess the direct effect by collecting the quantitative survey from the slum people. However, it has not been possible because of the wide-spreading pandemic situation due to the lack of data. In this context, the KIIs were conducted to summarize the findings as a proxy. Secondly, the estimation for indirect effect was not robust as most of the coefficients were statistically insignificant. Despite all these limitations, this study has explored some critical findings that could help policymakers. This study suggests the following recommendations;

(i) The government should provide good water and sanitation facilities to the slums so that their living condition improve. The implementation of this facility may reduce the government cost of free treatment in government hospitals.

(ii) In households where slum women work as a housemaid, they may be asked to contribute to the implementation cost of the water and sanitation facilities, and

(iii) Society can contribute to managing this facility as a social and moral obligation because of dependency on the slum dwellers for their services.

With the recommendations mentioned above and limitations, this study also recommends further robust studies on the similar topic so that both direct and indirect costs of water and sanitation-borne diseases could be identified and respective policies could be framed for the upliftment of the unprivileged people of the society who live in the slums of the city.